# Improving the Optical and Thermoelectric Properties of $Cs_2InAgCl_6$ with Substitutional Doping: A DFT Insight

K. C. Bhamu*[1,2], Enamul Haque[‡3], C.S. Praveen[4,5], Nandha Kumar[6], G. Yumnam[7], Md. Anwar Hossain[3], and Gautam Sharma[8]

[1]PMC Division, CSIR-National Chemical Laboratory, Pune 411008, Maharashtra, India
[2]Department of Physics, Gramin Mahila P. G. College, Sikar 332024, Rajasthan, India
[3]Department of Physics, Mawlana Bhashani Science and Technology University, Santosh, Tangail-1902, Bangladesh.
[4]International School of Photonics, Cochin University of Science and Technology, University Road, South Kalamassery, Kalamassery, Ernakulam, Kerala 682022, India
[5]Inter University Centre For Nano Materials and Devices, Cochin University of Science and Technology, University Road, South Kalamassery, Kalamassery, Ernakulam, Kerala 682022, India
[6]The Abdus Salam International Centre for Theoretical Physics, Strada Costiera 11, 34151 Trieste, Italy
[7]Department of Physics and Astronomy, University of Missouri, Columbia, USA 65211
[8]Department of Physics, Indian Institute of Science Education and Research, Pune 411008

*Corresponding author Email: kcbhamu85@gmail.com,
[‡]Email: enamul.phy15@yahoo.com

**Abstract**

New generation Indium based lead-free $Cs_2InAgCl_6$ is a promising halide material in photovoltaic applications due to its good air stability and non-toxic behaviour. But its wide band gap (> 3 eV) is not suitable for solar spectrum and hence reducing the photoelectronic efficiency for device applications. Here we report a significant band gap reduction from 3.3 eV to 0.65 eV by substitutional doping and its effect on opto-electronic and opto-thermoelectric properties from first-principles study. The results predict that Sn/Pb and Ga & Cu co-doping enhance the density of states significantly near the valence band maximum (VBM) and thus reduce the band gap by shifting the VBM upward while the alkali-metal (K/Rb) slightly increase the band gap. A strong absorption peak near Shockley–Queisser limit is observed in co-doped case while in Sn/Pb-doped case, we notice a peak in the middle of the visible region of solar spectrum. The nature of band gap is indirect with Cu-Ga/Pb/Sn doping with a significant reduction in band gap, from 2.85 eV to 0.65 eV in the case of Ga-Cu co-doping. We observe a significant increase in the power factor (PF) (2.03 mW/mK$^2$) for n-type carrier in Pb-doping, which is ~3.5 times higher than the pristine case (0.6 mW/mK$^2$) at 500 K.

[‡]Author contributed equally to the work.

## 1. Introduction

Expeditious increase in World's energy demands accelerated the quest for alternative fuels based on renewable energy sources and have been emerged as a global challenge [1]. The solar light is considered to be a viable alternative to traditional fossil fuels, as it can be regarded as a clean and inexhaustible source of energy, which effectively compromises the environment and the future. Large scale efforts have been steadily aggravated over the last few decades to develop environment-friendly and cost-effective energy technologies which utilize the solar spectrum using solar cells and waste heat due to Seebeck and Peltier effect [2]. Therefore, scientists have put an enormous effort in discovering efficient and low-cost materials which effectively utilize the UV-visible region of the solar spectrum.

Halide perovskites are being studied for the last few decades as a potential candidate for solar energy harvesting [3]. After the first successful attempt in the photovoltaic application by Kojima et al. [4], the organolead halide perovskites, $CH_3NH_3PbBr_3$, (OHP) have gained significant scientific attention and have been emerged as a star material in the solid-state solar cell with stability issue since 2012 [5]. Available featured experimental [6,7] and theoretical studies [8–11] on halide perovskites mainly focus on optoelectronic and thermoelectric applications [12–14]. OHP's are proposed as economical and efficient materials like Silicon and SnSe for the optoelectronic and thermoelectric application, respectively. Under the practical operating conditions, i.e. moisture, air, and temperature, the organometal halide perovskites decompose into secondary phases [15,16]. The long-term stability and toxicity of Pb still lags behind their high efficiency, and it remains as an ultimate challenge [17]. There have been efforts to replace the $CH_3NH_3$ with metal cations to improve the stability and Pb with non-toxic cations. Recently double perovskites emerged as a green alternative to the poisonous lead halide perovskite. They also offer viable engineering route to tune the band gap with various cation and anion choices [16,18–23]. A successful attempt was made for inorganic halide perovskites in 2016 where first double halide perovskite, $Cs_2AgBiX_6$ ($X$ = Br, Cl), was synthesized and tested for photovoltaic applications by Slavney et al. [20] and McClure *et al.* [22]. Later on, Xiao et al. [24] reported that the week coupling between 6s and 6p orbitals of Bi and Br, respectively makes $Cs_2AgBiX_6$ less stable than OHP. The bandgap of $Cs_2AgBiX_6$ is beyond the Shockley–Queisser limit required for a single-junction solar cell [25]. Theoretical studies of Han et al. [26] based on density functional theory (DFT) contradicts the experimental results (structural stability) reported by Slavney *et al.* [20] and McClure *et al.* [22], and predicts its decomposition into binary and ternary compounds above 700 K.

Volonakis et al. [27] reported $Cs_2InAgCl_6$ as a direct band gap (3.3 eV) material, which has air stability, non-toxicity, and Earth abundance. This double perovskite possesses a highly tuneable direct band gap in the visible range. The measured direct band gap of $Cs_2InAgCl_6$ is too far from the optimum value of Shockley–Queisser limit required for high-performance solar cells. Such a large band gap of $Cs_2InAgCl_6$ limits the effective carrier transport between valence band maximum and conduction band minimum thus limiting the efficiency of the thermoelectric devices [28]. However, it could be a suitable alternative of OHP if the band gap can be reduced by band gap engineering. Double perovskites have the peculiar feature of wide compositional tunability. They offer to alter the atomic composition by doping or alloying, which is a straightforward approach to engineer the band gap. Here we attempt to tune the band gap and hence optoelectronic and opto-thermoelectric properties of $Cs_2InAgCl_6$ by substitutional doping.

## 2. Crystal structure and Computational methods

The lead-free halide double perovskite $Cs_2InAgCl_6$ crystallize in the face-centred cubic structure with space group $Fm-3m$ (225) [27] as shown in the left panel of Fig. SI-1(supporting information). The Cs/In/Ag/Cl occupies 8c/4a/4b/24e Wyckoff positions. The optimized lattice parameters of $Cs_2InAgCl_6$ with other available data are listed in Table SI-1 for comparison and they are in agreement with previous reports [27]. To model the 25% substitutional doping at different atomic sites, two Cs-atoms located at (¾, ¾, ¾) and (¼, ¼, ¼) positions are substituted by Rb or K. In Pb and Sn-doped cases, In at (0, 0, 0) and Ag at (½, ½, ½) are substituted by Pb and Sn. For co-doping, we replaced the In at (0,0,0) by Ga and Ag at (½, ½, ½) by Cu resulting in 25% effective doping in the parent system. The doped system adopts different space group as mentioned in Table-S1. Further crystallographic details can be obtained in Ref. [27].

We have performed all the first principle calculations using full potential augmented plane wave method (FP-LAPW) as implemented in WIEN2k code [29]. We used the generalized gradient approximation (GGA) with Perdew-Burke-Ernzerhof (PBE) functional for the description of electron exchange and correlation [30,31]. To calculate the electronic properties with an improved band gap, we employed Becke and Johnson potential [32] and its modified versions [33,34]. Table-SI-1 depicts that the mBJ proposed by Koller et al. [35] (here after referred as KTB-mBJ) gives the better estimate of band gap than other version of mBJs. Recently, Jishi et al. [36] reported that, with appropriate tuning of parameter $c$ (Eqn. 1 in Ref [32]), can reproduce the experimental band gap of perovskites. But in the present investigation, we find that the band gap is overestimated compared to the experimental band gap, and hence we do not report the results calculated from Jishi's mBJ potential. Therefore, all the relevant properties of interest are computed with KTB-mBJ in this study. We set the size of k-mesh to be $12 \times 12 \times 12$, the product of smallest of all atomic sphere radii ($R_{mt}$) and the plane wave cut-off ($K_{max}$) 8.5, the magnitude of the largest vector ($G_{max}$) 12, $l_{vnsmax}$=6, energy, charge and force convergence criteria to be $10^{-5} Ry$, 0.0001e and 0.05 mRy/a.u., respectively. The self-consistent field calculations are performed at $10 \times 10 \times 10$ and $5 \times 5 \times 5$ k-mesh for pristine and doped cases with keeping all the other parameters as described above and mentioned in Table 1. For the computation of the electronic and optical properties, $14 \times 14 \times 14$ and $7 \times 7 \times 7$ shifted k-mesh grid was used for pristine and doped systems. We calculated optical properties by using the method described in reference [37]. Thermoelectric (TE) properties such as Seebeck coefficient (α), electrical conductivity (σ), electronic thermal conductivity ($k_e$), power factor (PF) and thermoelectric figure of merit (ZT), are computed using Boltzmann transport theory within constant relaxation time approximation (CRTA) and rigid band approximation (RBA) [28,38,39]. Convergence has been achieved in TE coefficients (change less than 1%) with a dense k-grid of 28 × 28 × 28 and 14 × 14 × 14 for the pristine and the doped cases, respectively.

## 3. Results and Discussions

### 3.1. *Electronic Structure*

Fig.1 (a-c) depicts the band structure of pristine $Cs_2InAgCl_6$ computed using PBE (Fig. 1a), PBE+SOC (Fig. 1b) and KTB-mBJ+SOC (Fig. 1c). The overall features of the band structure of $Cs_2InAgCl_6$ are similar to the previous reports as in Ref. [40]. It is to be noted that the band gap (2.85 eV) computed using TB-mBJ agrees well with the values (2.53 eV; TB-mBJ) reported by co-authors [40] but still underestimates the experimental band gap (3.3 eV) owing to the difference in the mBJ potentials used. Additionally, it is clear from Fig. 1(c) that KBT-mBJ+SOC ($E_g$=2.85 eV) has improved the PBE band gap (1.0 eV; Fig. 1a) and we observe that the inclusion of SOC (Fig. 1b) slightly changes the band dispersion in the valence band.

Further information about the density of states (DOS) and projected DOS for the studied systems are given in Figs. SI-(2-5).

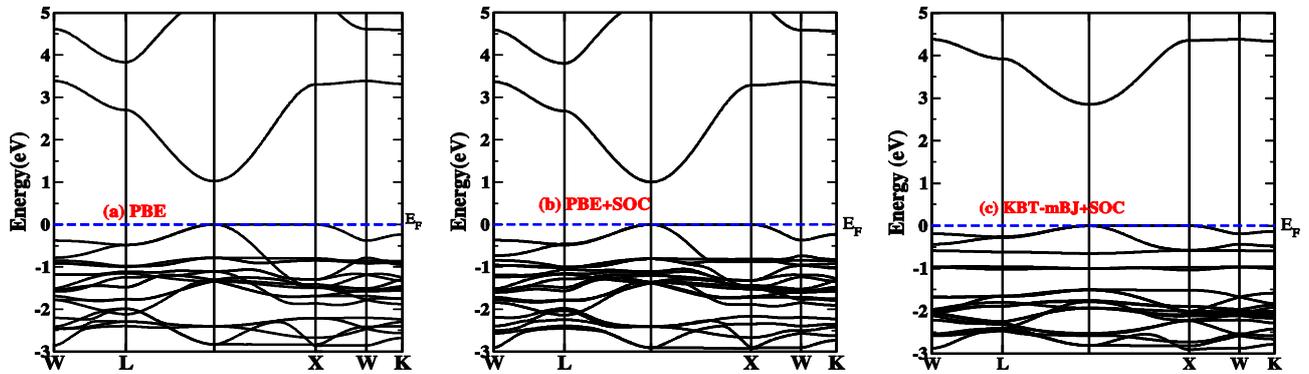

Fig. 1: Electronic dispersion relations of $Cs_2InAgCl_6$ using: (a) PBE (b) PBE+SOC, and (c) PBE with KTB-mBJ+SOC. The blue dash line represents the Fermi level ($E_F$).

To engineer the band gap, we have employed a bifold substitutional doping at the different atomic sites, i.e., isoelectronic (K/Rb at Cs site) and non-isoelectronic (Sn, Pb; Cu & Ga co-doping).

We observed for the isoelectronic doping that, for the fully relaxed system, lattice parameters are found to change by less than 0.5% compared to pristine case. Since K/Rb is doped in cage of the octahedra's (Cs-site) of AgCl$_6$ and InCl$_6$ thus their valence electron (ns$^1$) is not expected to contribute near VB/CB edges because it is located at the deep valence states near ~ 6 eV. The same can be confirmed from electronic structure calculations, shown in Fig. SI-(3, 4). The band gap increases by 0.18 (0.16 eV) with K (Rb) doping in comparison to the pristine case which is resulted due to the size effect of the dopant [41] and the band dispersion remain intact.

It is reported [42] that the band dispersion near Fermi level can be altered significantly by alloying the octahedra cation site (Ag/In) to enhance the opto-electronic and opto-thermoelectronic properties of the material. Therefore we opt Ag/In site for the non-isoelectronic doping.

Fig. 2(a-b) are representing Pb and Sn doped band structure, respectively. Our findings reveal that Sn is better dopant candidate as compared to Pb for optoelectronic applications. The detailed description can be found in the next section.

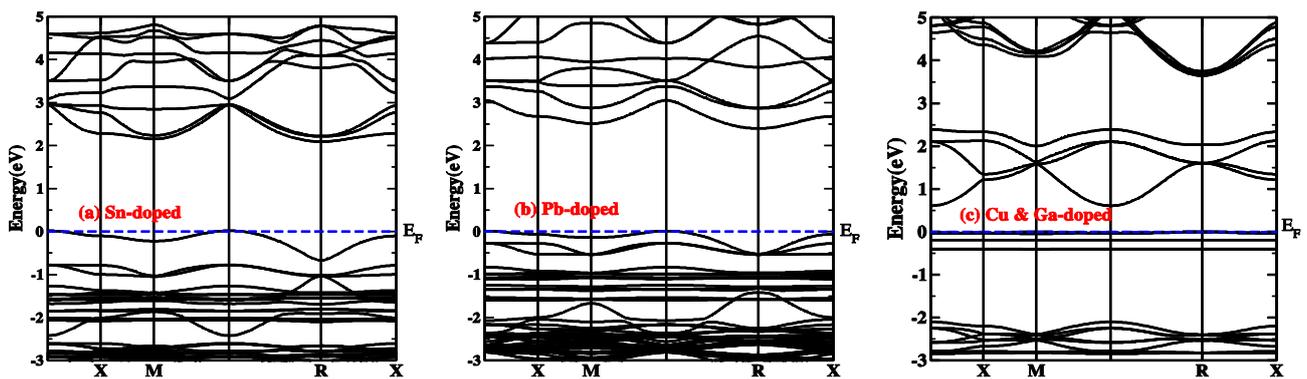

Fig. 2: Electronic dispersion relations of: (a) 25% Sn-doped, (b) 25% Pb-doped, and (c) 12.5% each Cu & Ga-doped Cs$_2$InAgCl$_6$ by using PBE functional with KTB-mBJ potential including SO effect. The blue dash line represents the Fermi level.

The doped system becomes indirect semiconductor with the band gap of 2.40 (2.08) eV with Pb (Sn) doping. Reduction in the band gap with Pb relative to Sn doping can be understood from the location of In-s states which gets shifted by ~0.3 eV when system is doped with Pb relative to Sn. The VBM occurs at Γ point whereas in the conduction band there are two minima present at M and R. Both the minima in the conduction band are separated by ~0.07 eV which give rise the valley degeneracy responsible for high figure of merit [ZT] for n-type carriers [43].

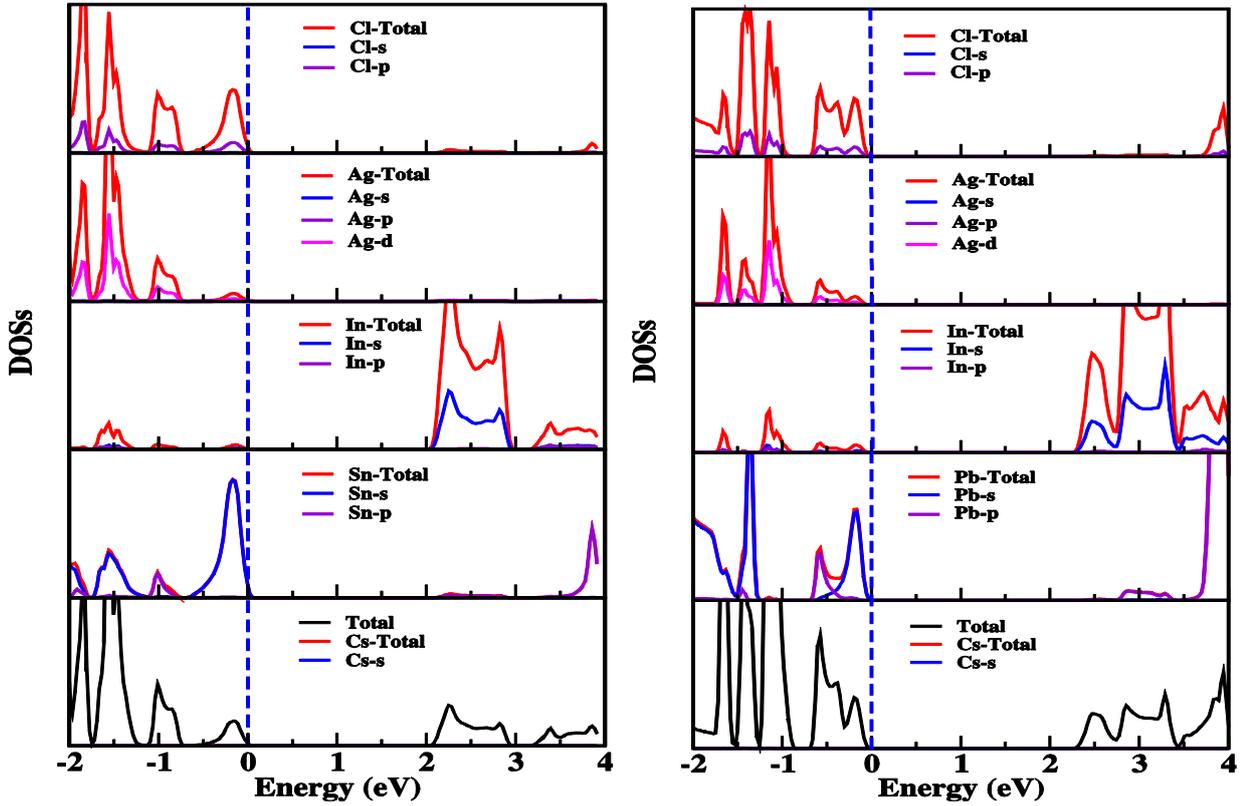

Fig, 3. Calculated total, atom, and orbital decomposed density of states of Sn- (Left panel) and Pb-doped (Right panel) Cs2InAgCl6 by using PBE functional with KTB-mBJ potential.

Interestingly, we find the presence of Sn/Pb-*s* states in proximity of Fermi level giving rise to increased charge carriers which is advantageous especially when it comes to thermoelectrics.

Fig. 2(c) illustrates the band structure of CGCD. The top of the valence band is mainly composed of Cu-*d* state with slight contribution from In-*s/p* states. In the conduction band, the In-*s* state is shifted towards the Fermi level by ~ 2.2 eV reducing the band gap value 0.65 eV which is lowest band gap found among various systems in the present study. The band gap is indirect in nature that prevents the fast recombination of electron-hole giving high exciton life time.

The calculated total density of states of the studied alloys is illustrated in Fig. SI-(2, 4, 5). In the next sections, we focus on the effect of reduction of the band gap on optical and thermoelectric properties.

## 3.2. Optical properties

After calculating Kohn-Sham (KS) electronic states at the k-grids mentioned in the section 2, we compute the optical absorption spectra using KTB-mBJ+SOC shown in Fig. 4 for pristine $Cs_2InAgCl_6$ along with its doped derivatives. Full description of the method and equations used in computing the optical properties is adopted from Ref. [37]. The change in the band gap by substitutional dopant clearly reflected in the optical spectrum. For photovoltaic applications, a material must show high optical absorption spectra in the visible region of the solar spectrum [44,45]. The optical absorption edge is calculated by extrapolation at ~3.6 eV (not shown in Fig.4), which gradually shifts towards lower energy and for CCGD system, the shift is drastic.

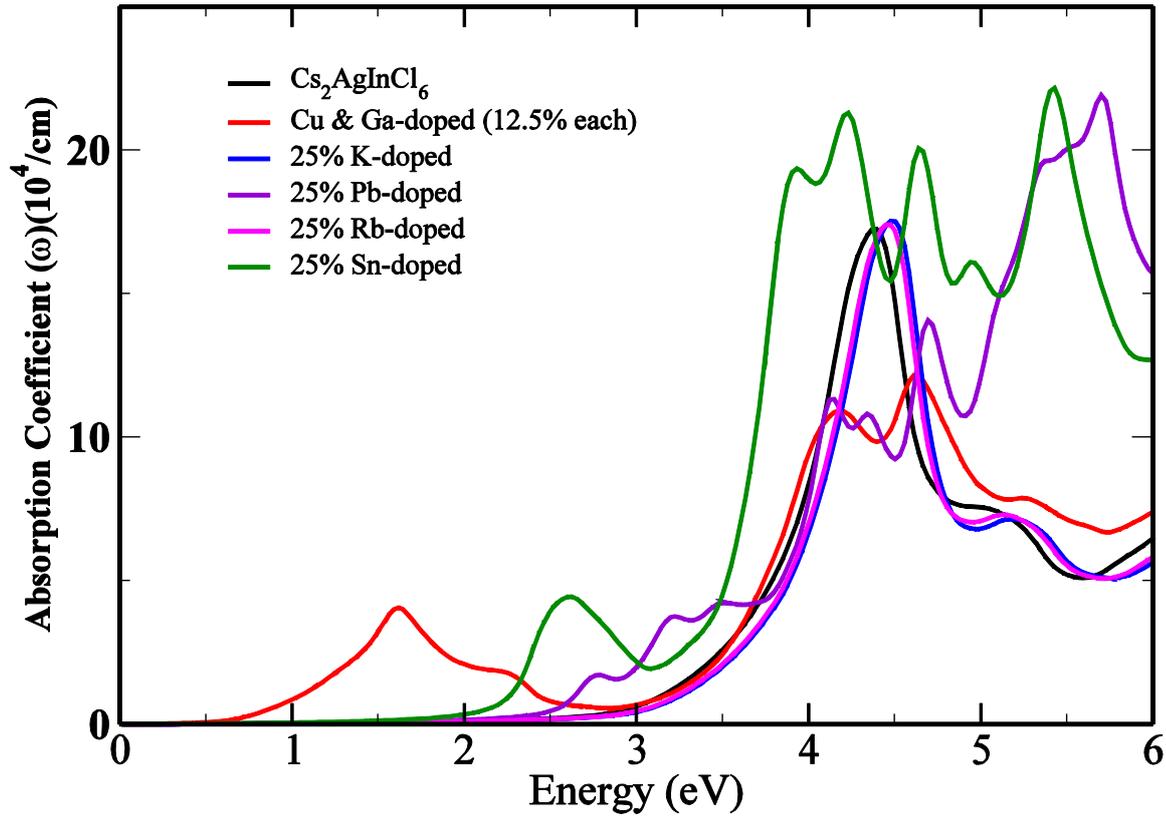

Fig. 4: Absorption coefficient of $Cs_2InAgCl_6$ along with doped cases as a function of photon energy.

Our results show that doping with suitable candidate can significantly improve the optical absorption starting from ultra-violet (UV) to IR (infra-red). The optical absorption strength for co-doped system is stronger than other doped systems towards the IR solar spectrum while in visible region Sn doped system is dominant. The present findings may serve as a guide for experimental studies on these non-isoelectronic doped systems.

### 3.3. *Thermoelectric properties*

The thermoelectric (TE) transport properties calculated using the linearized Boltzmann transport equation (BTE), as implemented in BoltzTraP code [38] which solves BTE within the constant relaxation time approximation (CRTA). BoltzTraP code interpolates the bands structure obtained from the DFT calculations, do the necessary integrations (Fermi integrals) at different temperature and Fermi level and provides required transport coefficients as a function of temperature and carrier concentration (for both p-type and n-type) shown in Figs. (5-7).

The performance of thermoelectric material is governed by the dimensionless figure of merit (ZT), given by [46,47]

$$ZT = \frac{\alpha^2 \sigma}{\kappa} T$$

where $\sigma$ is electrical conductivity, $\alpha^2\sigma$ is the power factor, T is absolute temperature, $\sigma$ is the electrical conductivity and $\kappa$ is thermal conductivity composed of electronic ($\kappa_e$) and lattice part ($\kappa_l$). We do not consider lattice part of the thermal conductivity here.

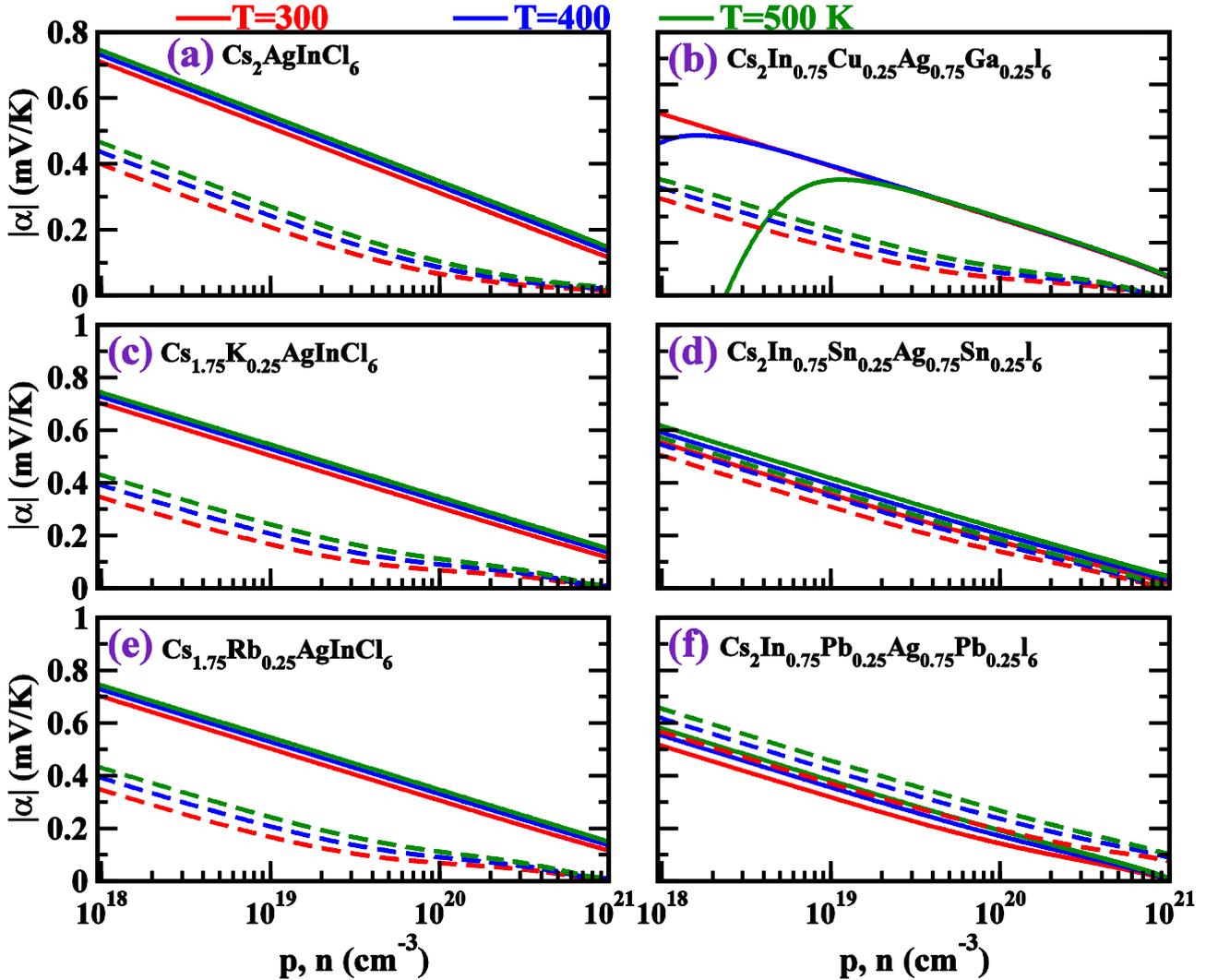

Fig. 5: The carrier concentration dependent Seebeck coefficient (|α|) at different temperatures for both *p*-type (solid) and *n*-type (dotted) of (a) Cs$_2$AgInCl$_6$, (b) Cu & Ga co-doped, (c) K-doped, (d) Pb-doped, (e) Rb-doped, and (f) Sn-doped.

Tuning the band gap by substitutional doping and thus band gap has changed the dispersion of electronic bands around Fermi level. This in turn governs the behaviour of Seebeck coefficient and electronic thermal conductivity and hence power factor.

Fig. 5 displays the Seebeck coefficient (|α|) of pristine and doped systems for p-type and n-type. The isoelectronic doping (with K and Rb) (Fig. 5: c, e) does not alter the |α| as there is no change in dispersion of bands. We find |α| shows the typical behaviour of decreasing the carrier concentration

and also computed |α| for hole is higher than for the electrons. Since |α| is directly proportional to effective mass of the charge carriers, therefore, with flat band dispersion (Γ-X) at valence band edges gives rise to large hole effective mass ($m^*_h$). On the other hand, we have parabolic band dispersion for conduction bands causing lighter electron effective mass ($m^*_e$). For Pb/Sn doped (Fig. 5d) system, the behaviour of |α| can be interpreted based on the band gaps and bands dispersion. Notice for all temperatures, |α| of electrons with Pb-doped system is ~3 times higher that observed for pristine case at n = $10^{20}$ cm$^{-3}$ which is resulted due to downward shift of In-s state. It is found that the Pb-doping is more favorable than pristine. Moreover, we find that |α| for electrons exceeds that for holes at higher carrier concentration and effect of which is reflected in power factor (PF). In case of CGCD, the narrow band gap triggers bipolar conduction at lower carrier concentrations and due to flat band the electrical conductivity is too low relative to other cases.

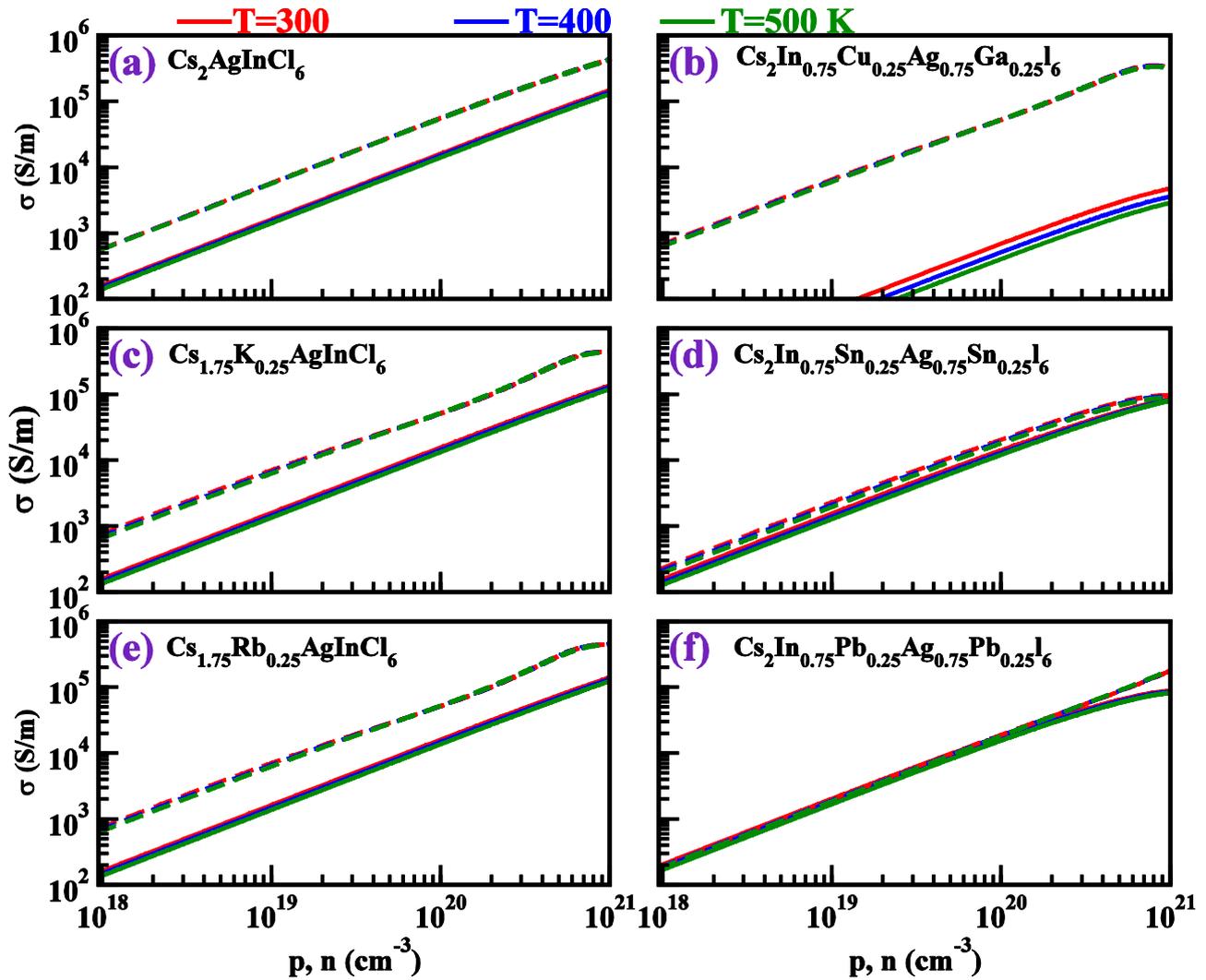

Fig. 6: Description is same as mentioned for Fig. 4 except the property which is electrical conductivity.

For all the cases, we obtain the electronic conductivity (σ) and PF by multiplying by τ = $10^{-14}$ s and the same approach is also adopted by others [38,48], and are displayed in Figs. (6-7). We see that σ increases with the increase of carrier concentration. In all the cases, σ is found to be more for electrons than holes due to parabolic dispersion of bands. However, in the Pb-doped system, σ for electrons and holes are found to be of similar order which is advantageous to boost performance with n-type doping.

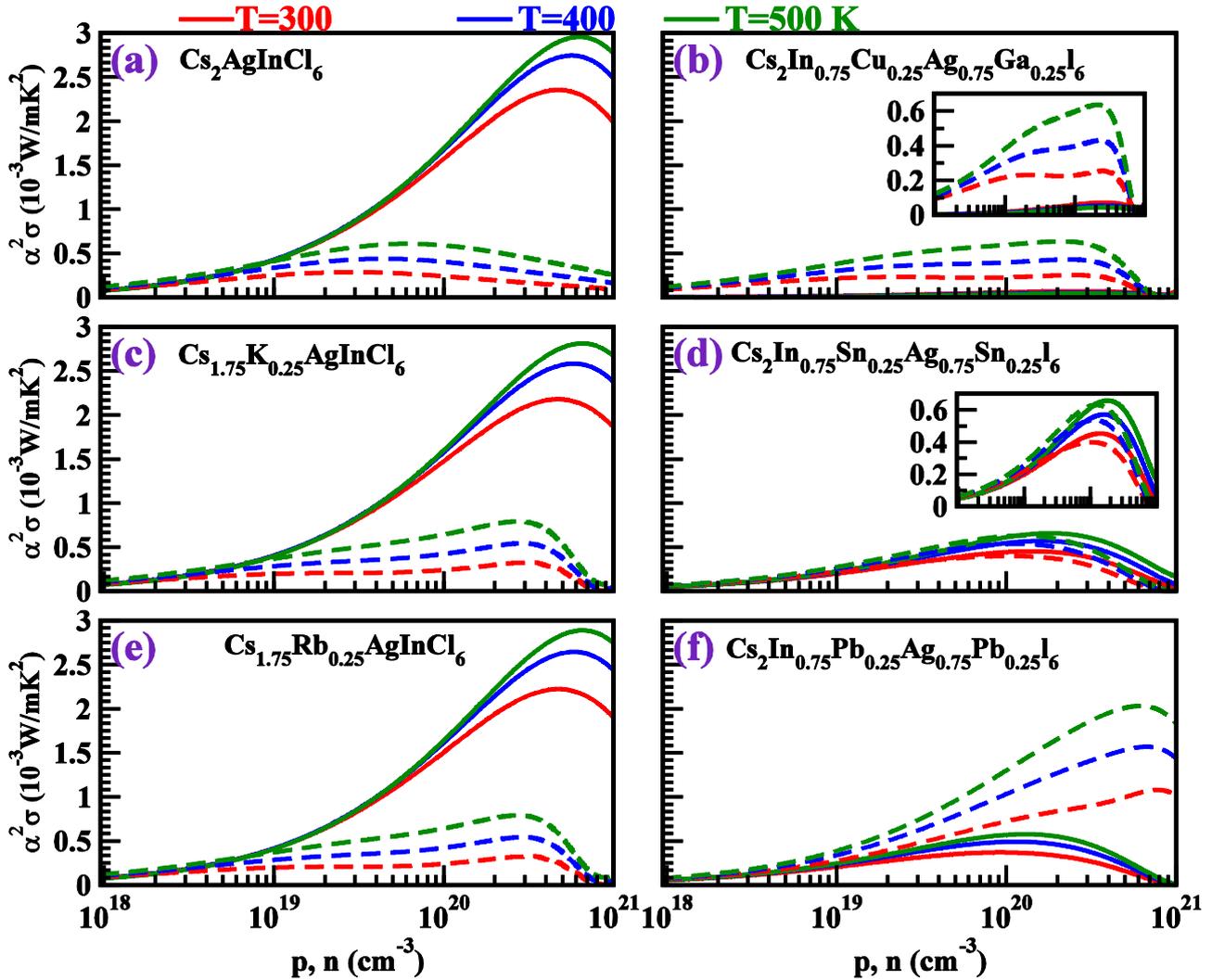

Fig. 7: Description is same as mentioned for Fig. 5 except the property which is power factor.

Temperature-dependent power factor which is product of $\alpha^2$ and $\sigma$ and is plotted as a function of carrier concentration in Fig. 7. Due to isoelectronic dopant (Fig. 7c, e), the behaviour of power factor for K-/Rb-doped case is nearly same to pristine case with p-type of doping and it is slightly higher than the pristine case for n-type doping. For the co-doped case (Fig. 7b) the optimized value of power factor with n-type doping is 0.6 mW/mK$^2$. The remarkable feature from our study is the significantly enhanced PF (2.03 mW/mK$^2$) for n-type carrier concentration for Pb-doped case where the PF is ~3.5 times higher than the pristine case (0.6 mW/mK$^2$) at 500 K.

## 4. Conclusions

In summary, we have studied the effect of K, Rb, and Cu-Ga/Pb/Sn substitution on electronic, optical and thermoelectric transport properties of $Cs_2InAgI_6$ using DFT based first-principles calculations. We have found that K/Rb-substitution increases the band gap slightly while Cu-Ga/Pb/Sn-substitution decreases the band gap significantly. K/Rb based studied system have been found to be a direct band gap semiconductors while indirect band gap nature is observed for other doped cases. The absorption peak is shifted to the visible spectrum from ultra-violet region for indirect band gap compounds but remains intact for direct band gap compounds. The optical absorption strength for co-doped system is stronger than other doped systems towards the IR solar spectrum while in visible region Sn doped system shows dominant absorption. The significant changes in thermoelectric properties has noticed for Pb- and co-doped cases. The flat band at valence band maximum is responsible for poor thermoelectric performance for co-doped system whereas the downward shift of In-s state in Pb-doped system enhances the PF by ~3.5 times higher than the pristine case (0.6 mW/mK$^2$) at 500 K with n-type doping and screens it out as a promising candidate for thermoelectric performance based on our investigations.

**Acknowledgements**

KCB and CSP acknowledge DST and DST-SERB for the SERB-NPDF and INSPIRE Faculty fellowship with award number IFA-18 PH217 and PDF/ 2017/002876, respectively.

The research used resources at the National Energy Research Scientific Computing Center (NERSC), which is supported by the Office of Science of the USDOE, United States under Contract No. DE-AC02- 05CH11231. High performance computational facility from Centre for Development of Advanced Computing, Pune also used for the present investigations.

# Supporting Information
# Improving the Optical and Thermoelectric Properties of $Cs_2InAgCl_6$ with Substitutional Doping: A DFT Insight

K. C. Bhamu*[1,2], Enamul Haque[‡3], C.S. Praveen[4,5], Nandha Kumar[6], G. Yumnam[7], Md. Anwar Hossain[3], and Gautam Sharma[8]

[1]PMC Division, CSIR-National Chemical Laboratory, Pune 411008, Maharashtra, India
[2]Department of Physics, Gramin Mahila P. G. College, Sikar 332024, Rajasthan, India
[3]Department of Physics, Mawlana Bhashani Science and Technology University, Santosh, Tangail-1902, Bangladesh.
[4]International School of Photonics, Cochin University of Science and Technology, University Road, South Kalamassery, Kalamassery, Ernakulam, Kerala 682022, India
[5]Inter University Centre For Nano Materials and Devices, Cochin University of Science and Technology, University Road, South Kalamassery, Kalamassery, Ernakulam, Kerala 682022, India
[6]The Abdus Salam International Centre for Theoretical Physics, Strada Costiera 11, 34151 Trieste, Italy
[7]Department of Physics and Astronomy, University of Missouri, Columbia, USA 65211
[8]Department of Physics, Indian Institute of Science Education and Research, Pune 411008

*Corresponding author Email: kcbhamu85@gmail.com,
[‡]Email: enamul.phy15@yahoo.com

## SI 1. Structural details

The double-perovskite $Cs_2InAgCl_6$ crystallizes in a face-centred cubic structure with 40 atoms per unit cell. The unit cell contains two octahedra: one $InCl_6$ and other $AgCl_6$. These two octahedra alternate along the different crystallographic planes: [100], [010], and [001].

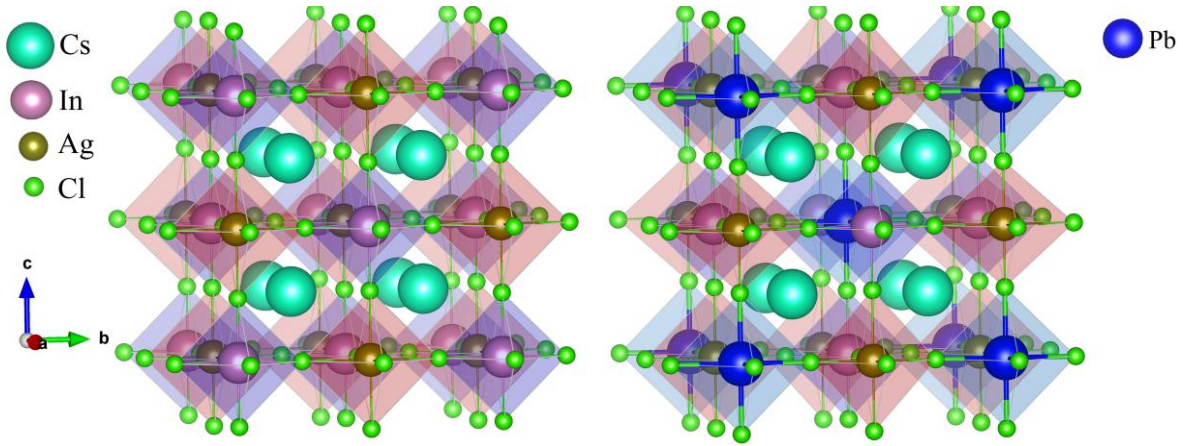

Fig. SI-1: Crystal structure of double perovskite $Cs_2InAgCl_6$ (left panel),and Cu-Ga/Pb/Sn-substituted-$Cs_2InAgCl_6$ (right panel).

The unit cell of Cs2InAgCl6 is shown in the left panel of Fig. SI-1. The K/Rb based structure is almost identical to the pristine $Cs_2InAgCl_6$ structure. The right panel shows the crystal structure of 25% Sn, Pb and 12.5% each Cu & Ga- doped $Cs_2InAgCl_6$ crystal structure. This structure seems to be octahdrally distorted antiperovskite type structure.

**SI 2. Electronic structure**

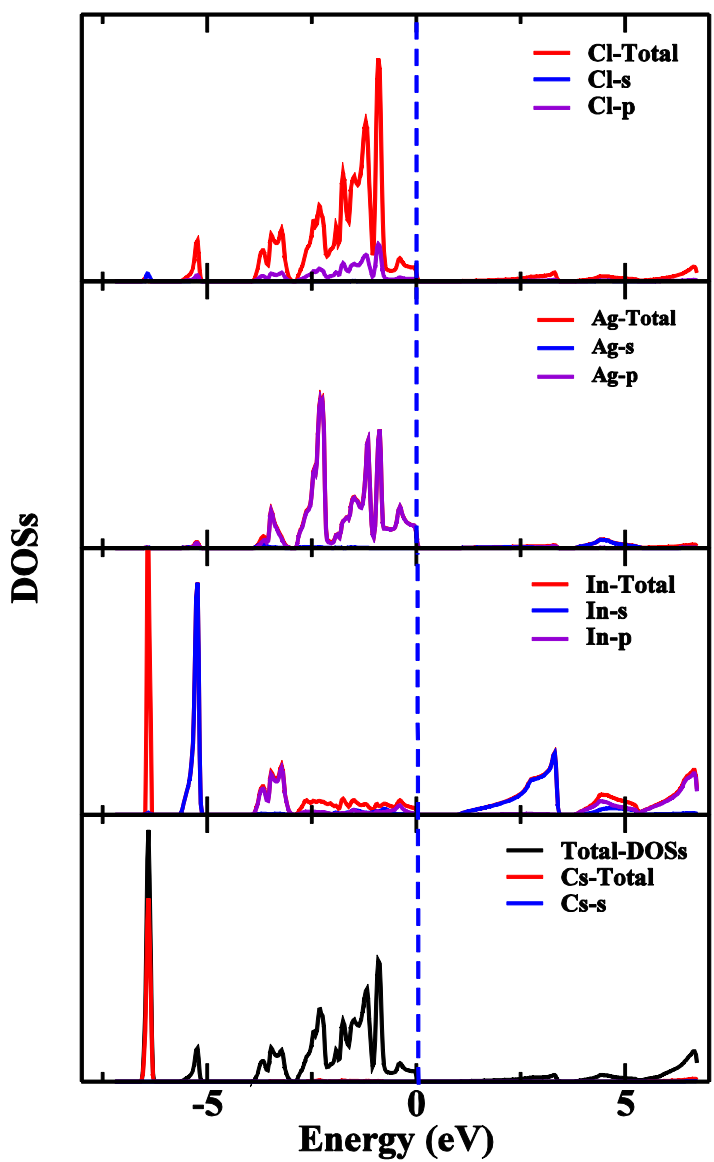

*Fig. SI-2: Calculated total, atom and orbital decomposed density of states of presitne Cs$_2$InAgCl$_6$ by using PBE functional with KTB-mBJ potential. The DOS including SO effect is shown by dash line.*

The other change observed with employing KTB-mBJ+SOC is that the density of states become sharper the increase of intensity in the entire range of the DOSs, Fig. SI- 2(a-c). We see from DOSs that in the pristine case the valence band maximum (VBM) is composed of Ag-$d$ and Cl-$p$ states while conduction band minimum is originated from In-$s$ and Ag-$s$ states. Assuming the similar trend for

PBE and PBE+SOC for doped materials as we observe for pristine, we report only KTB-mBJ+SOC band structure only for the doped case.

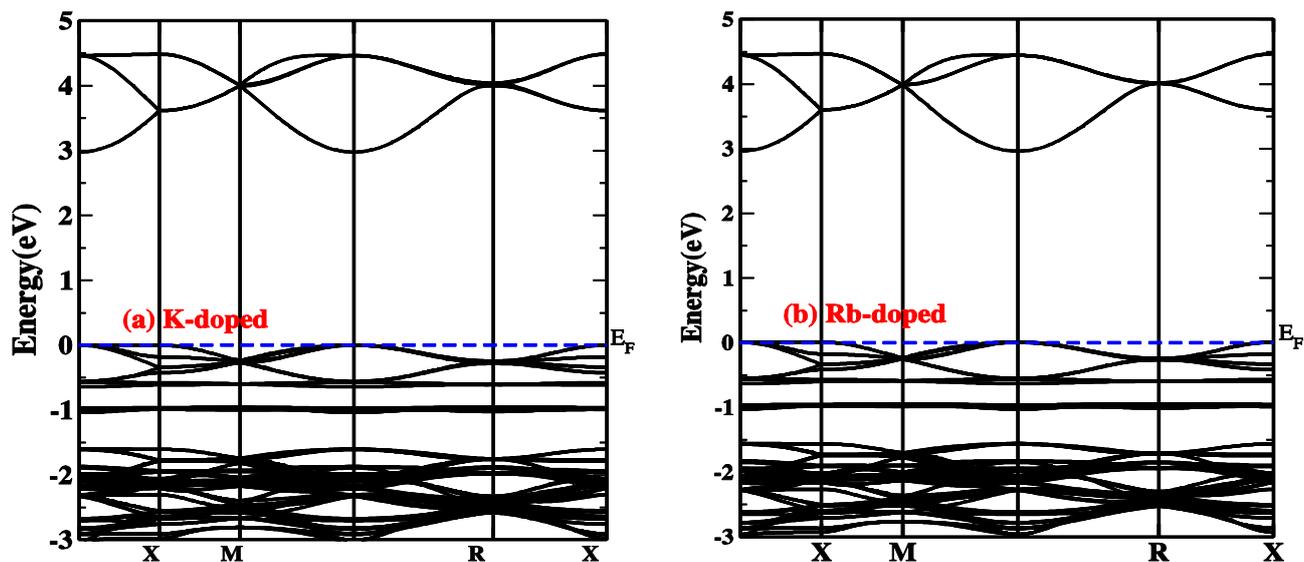

Fig. SI-3: Electronic dispersion relations of: (a) K- doped and (b) Rb-doped $Cs_2InAgCl_6$ by using PBE functional with KTB-mBJ potential including SO effect. The blue dash line represents the Fermi level.

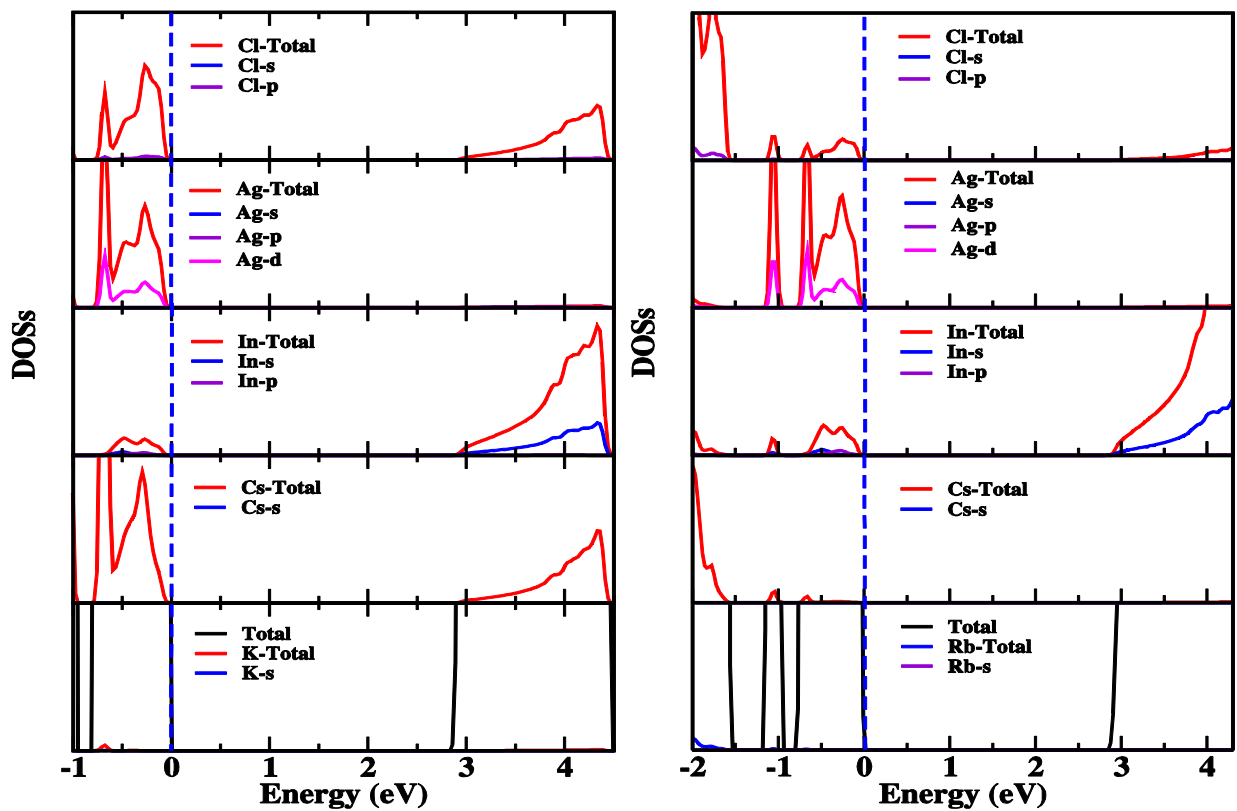

Fig. SI-4: Calculated total, atom, and orbital decomposed density of states of 25% K- (Left panel) and Rb-doped (Right panel).

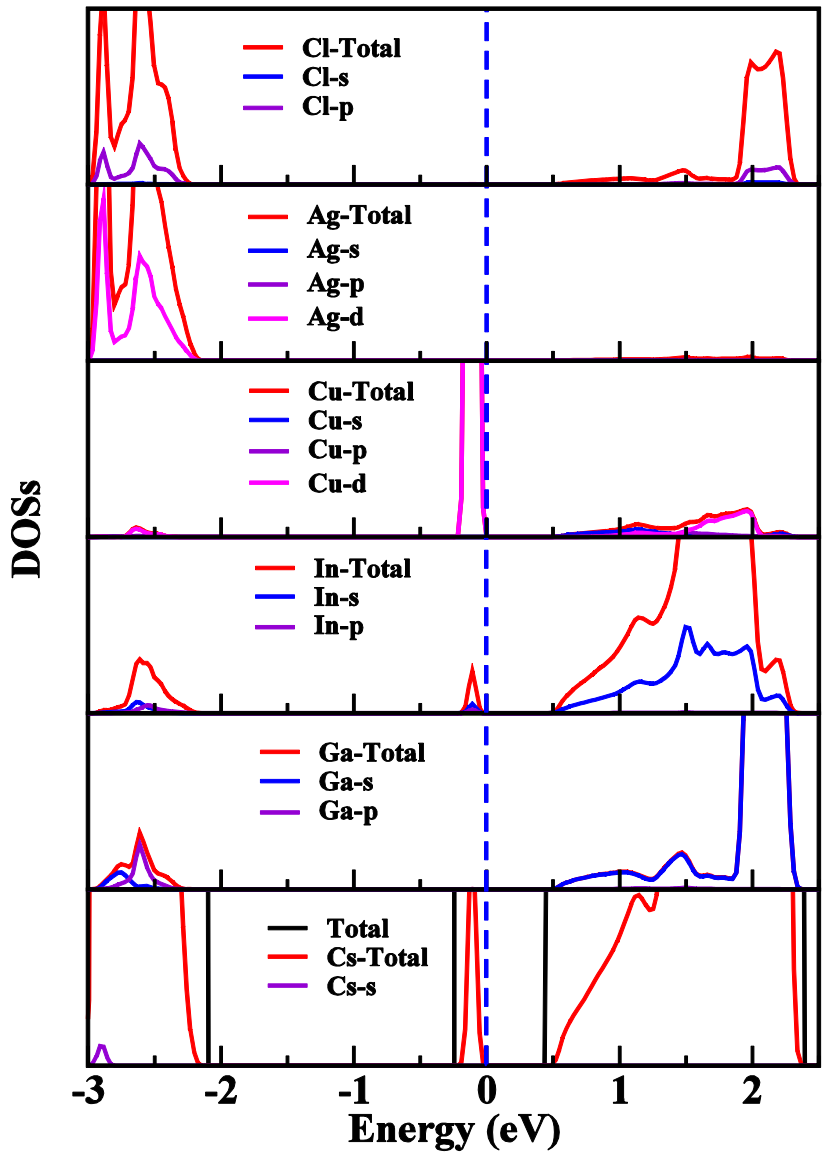

*Fig. SI-5: Calculated total, atom, and orbital decomposed density of states of 12.5% Cu & 12.5% Ga-co-doped $Cs_2InAgCl_6$ by using PBE functional with KTB-mBJ potential. The DOS including SO effect is not shown here.*

Table-SI-1: Equilibrium lattice parameters of pristine $Cs_2InAgCl_6$ along with its doped derivatives with experimental and others available data (symbols have their usual meaning).

| Material | Space group | a (Å) (PBE) | (Δa/a)% | Exp. [17] | Cal. (Others) |
|---|---|---|---|---|---|
| $Cs_2InAgCl_6$ | 225 (Fm-3m) | 10.676 | - | 10.467 | 10.20 (LDA) [17], 10.23 (LDA) [36], 10.62 (HSE) [17] |
| Cu-Ga co-doped | 221 (Pm-3m) | 10.545 | 1.23 | - | - |
| K-0.25 | 224 (Pn-3m) | 10.629 | 0.28 | - | - |
| Rb-0.25 | 224 (Pn-3m) | 10.646 | 0.44 | - | - |
| Sn-0.25 | 221 (Pm-3m) | 10.854 | 1.67 | - | - |
| Pb-0.25 | 221 (Pm-3m) | 10.916 | 2.25 | - | - |

Pb/Rb/K/Cs/Sn/In/Ag*/Cl*: 2.5/2.5/2.5/2.5/2.5/2.37/2.4/1.85
*$R_{mt}$ for Ag/Cl for pristine case is 2.5/2.04.

Table-SI-2: Calculated band gap of $Cs_2InAgCl_6$ by using different approaches of functional along with other theoretical and experimental value. The band gap of the studied alloys is listed for PBE functional and KTB-mBJ potential only.

| Compound | PBE | | TB-mBJ | | nKTB-mBJ | | KTB-mBJ | | HSE | PBE0 | EXP. |
|---|---|---|---|---|---|---|---|---|---|---|---|
| | PBE | +SOC | mBJ | +SOC | mBJ | +SOC | mBJ | +SOC | | | |
| $Cs_2InAgCl_6$ | 1.03 | 1.004 | 2.57 | 2.46 | 2.68 | 2.56 | 2.91 | 2.85 (d) | 2.6[1] | 2.7[1] | 3.3[1] |
| Cu &Ga | 0.12 | 0.09 | - | - | - | - | - | 0.65 (I) | - | - | - |
| K-0.25 | 1.11 | 1.08 | - | - | - | - | - | 3.03 (d) | - | - | - |
| Rb-0.25 | 1.11 | 1.09 | - | - | - | - | - | 3.01 (d) | - | - | - |
| Pb-0.25 | 2.13 | 1.85 | - | - | - | - | - | 2.40 (I) | - | - | - |
| Sn-0.25 | 1.66 | 1.59 | - | - | - | - | - | 2.08 (I) | - | - | - |